\documentclass[aps,prb,amsmath,amssymb,amsfont,showpacs,superscriptaddress,reprint]{revtex4-1}

\usepackage{amsmath}
\usepackage{xcolor}
\usepackage{graphicx}
\usepackage{times}
\usepackage{natbib}

\begin{document}

\title{Transport, Magnetic and Vibrational Properties of Chemically Exfoliated Few Layer Graphene}

\author{Bence G. M\'{a}rkus}
\affiliation{Department of Physics, Budapest University of Technology and Economics, POBox 91, H-1521 Budapest, Hungary}

\author{Ferenc Simon}
\affiliation{Department of Physics, Budapest University of Technology and Economics, POBox 91, H-1521 Budapest, Hungary}

\author{Julio C. Chac\'{o}n-Torres}
\affiliation{Institute of Experimental Physics, Freie Universit\"at Berlin, Arnimallee 14, 14195 Berlin, Germany}

\author{Stephanie Reich}
\affiliation{Institute of Experimental Physics, Freie Universit\"at Berlin, Arnimallee 14, 14195 Berlin, Germany}

\author{P\'{e}ter Szirmai}
\affiliation{Institute of Physics of Complex Matter, FBS Swiss Federal Institute of Technology (EPFL), CH-1015 Lausanne, Switzerland}

\author{B\'alint N\'afr\'adi}
\affiliation{Institute of Physics of Complex Matter, FBS Swiss Federal Institute of Technology (EPFL), CH-1015 Lausanne, Switzerland}

\author{L\'{a}szl\'{o} Forr\'{o}}
\affiliation{Institute of Physics of Complex Matter, FBS Swiss Federal Institute of Technology (EPFL), CH-1015 Lausanne, Switzerland}

\author{Thomas Pichler}
\affiliation{Faculty of Physics, University of Vienna, Strudlhofgasse 4, A-1090 Vienna, Austria}

\author{Philipp Vecera}
\affiliation{Department of Chemistry and Pharmacy and Institute of Advanced Materials and Processes (ZMP), University of Erlangen-Nuremberg, Henkestrasse 42, 91054 Erlangen, Germany}

\author{Frank Hauke}
\affiliation{Department of Chemistry and Pharmacy and Institute of Advanced Materials and Processes (ZMP), University of Erlangen-Nuremberg, Henkestrasse 42, 91054 Erlangen, Germany}

\author{Andreas Hirsch}
\affiliation{Department of Chemistry and Pharmacy and Institute of Advanced Materials and Processes (ZMP), University of Erlangen-Nuremberg, Henkestrasse 42, 91054 Erlangen, Germany}


\begin{abstract}
We study the vibrational, magnetic and transport properties of Few Layer Graphene (FLG) using Raman and electron spin resonance spectroscopy and microwave conductivity measurements. FLG samples were produced using wet chemical exfoliation with different post-processing, namely ultrasound treatment, shear mixing, and magnetic stirring. Raman spectroscopy shows a low intensity D mode which attests a high sample quality. The G mode is present at $1580$ cm$^{-1}$ as expected for graphene. The 2D mode consists of 2 components with varying intensities among the different samples. This is assigned to the presence of single and few layer graphene in the samples. ESR spectroscopy shows a main line in all types of materials with a width of about $1$ mT and and a $g$-factor in the range of $2.005-2.010$. Paramagnetic defect centers with a uniaxial $g$-factor anisotropy are identified, which shows that these are related to the local sp$^2$ bonds of the material. All kinds of investigated FLGs have a temperature dependent resistance which is compatible with a small gap semiconductor. The difference in resistance is related to the different grain size of the samples.
\end{abstract}

\maketitle

\section{Introduction}

Novel carbon allotropes gave an enormous boost to condensed-matter and molecular physics at the end of the last century. The process was started with the discovery of fullerenes \cite{kroto1985} and carbon nanotubes \cite{iijima1991}, but for the biggest breakthrough we had to wait until 2004 \cite{novoselov2004}.
Since its discovery graphene became one of the most important materials in condensed-matter physics. Being the basis of all other novel carbon allotropes \cite{geim2007,geim2009} (fullerenes, nanotubes, graphite), understanding graphene is crucial. The mechanical and electronic properties of graphene such as high fracture strength, high elasticity, low resistance, high carrier mobility, quantum Hall-effect make it an outstanding material for diverse applications \cite{neto2009}. However, one of the remaining obstacles for the applicability of graphene is mass production with controlled quality and graphene layer size.

High quality graphene can be prepared by mechanical exfoliation (also referred as mechanical cleavage), but only in small amounts on various substrates (maximum available size is still in the scale of microns \cite{Jayasena2011}). Epitaxial growth of graphene on various substrates \cite{Berger2006,Zhou2007,Sutter2008} is an alternative but the up-scalability of this method is limited and the resulting sample qualities needs yet to be improved. On the other hand, with chemical vapor deposition (CVD) high yields are achievable \cite{Obraztsov2007,Malesevic2008,Kim2009,Obraztsov2009,Reina2009,Mattevi2011,Yu2011} in a poorer quality due to the enormous number of defects. Another problem with the CVD method is that it still requires a substrate. Being a material of an atomically thin layer on a substrate is a serious issue when one would like to apply bulk characterization methods, such as Electron Spin Resonance spectroscopy (ESR) or macroscopic transport measurements (e.g. microwave conductivity). The substrate also has a negative effect on the electronic and vibrational properties of graphene (e.g. electronic interactions, and induced strain). These effects are visible when one tries to compare the results of free-standing graphene \cite{Chen2009} with graphene on other substrates: Si-SiO$_2$ \cite{Ferrari2006}, Si-SiO$_2$ and ITO \cite{Das2008}, SiC \cite{Faugeras2008}, glass \cite{Tsurumi2013}.

Other ways to create graphene in a mass production is reduction from graphite/graphene oxide (GO) and wet chemical exfoliation from graphite intercalation compounds (GICs) with various solvents. Reduction process is feasible in many chemical and biological routes with different quality of the final product \cite{Navarro2007,Osvath2007,Stankovich2007,Eda2008,Park2009,Dreyer2010,Shao2010,Zhang2010,Chen2010,Chen2010-2,Pei2010,Salas2010,Pei2012}. In general, the quality of final product may vary in a large scale but always contains residual oxygen, missing carbon atoms, free radicals, and dangling bonds therefore one can end up with a thermally metastable material \cite{Eigler2012,Eigler2013,Eigler2013-2,Eigler2014-2,Eigler2014}. 

Wet chemical or liquid phase exfoliation is the most promising way to mass produce high quality materials without disturbing the effects of the substrate \cite{Hernandez2008,Valles2008,Englert2009,Lotya2009,Englert2011,Catheline2011,Englert2012,Catheline2012,Kuila2012,Coleman2013,Hirsch2013,Vecera2014}. For the optimal quality of the outcome the effect of solvent \cite{Vecera2015} and the mechanical post-procession has to be examined. Here we report the transport, magnetic and vibrational properties of Wet Chemically Exfoliated (WCEG) Few Layer Graphene (FLG) using microwave conductivity, electron spin resonance and Raman spectroscopies.

\section{Experimental}

We studied three WCEG species which were prepared by different mechanical routes: ultrasounded (US), shear mixed (SM) and stirred (ST). All kinds were produced from saturate intercalated potassium graphite powder, KC$_8$ using DMSO solvent for wet exfoliation (full protocol is described in Ref. \cite{Vecera2014}). The starting material, SGN18 graphite powder (Future Carbon) and Grade I bulk HOPG (SPI) were taken into comparison. Mechanical post-processing were ultrasound treatment, shear mixing and magnetic stirring. The procedure was done under argon atmosphere. The pristine materials were cleaned under high vacuum ($10^{-7}$ mbar) at $400^{\circ}$C for one hour to get rid of the remaining solvent and impurities. Raman measurements were carried out in a high sensitivity single monochromator LabRam spectrometer \cite{Fabian2011} using $514$ nm laser excitation, $50\times$ objective with $0.5$ mW laser power. For ESR measurements a Bruker Elexsys E580 X-band spectrometer was used. Microwave conductivity measurements were done with the cavity perturbation technique \cite{Klein1993,Donovan1993} extended with an AFC feedback loop to increase precision \cite{Nebendahl2001}. The photographs were taken with a Nikon Eclipse LV150N optical microscope using $5\times$ (for FLG) and $10\times$ (for SGN18) objectives.

\section{Results and Discussion}

To get an insight on which mechanical post production method produces the best quality, the vibrational, electronic and transport properties of the materials have to be investigated. We discuss the Raman, ESR and microwave conductivity results.

\subsection{Raman spectroscopy}

\begin{figure}[h!]
\includegraphics[width=\linewidth]{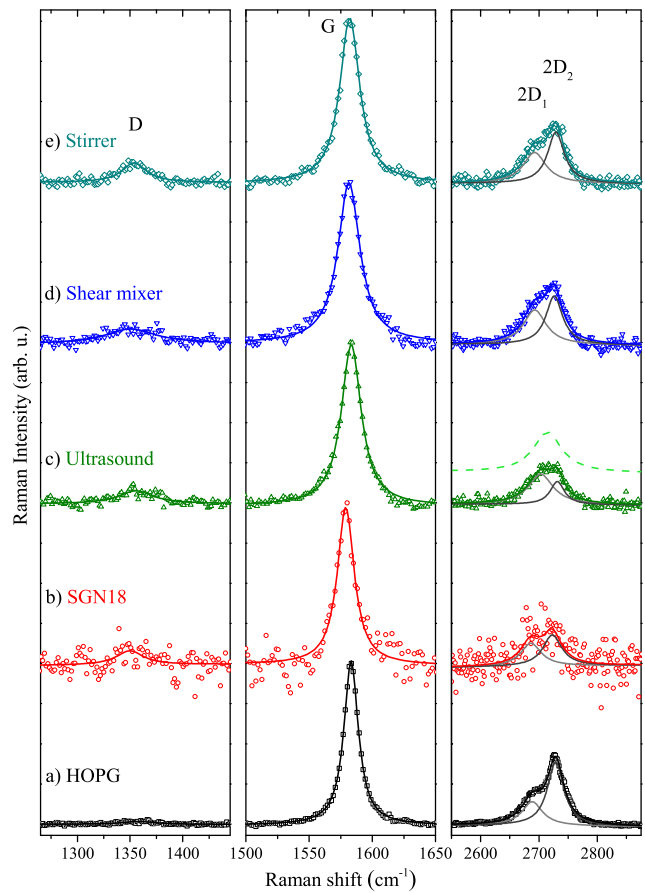}
\caption{D, G, 2D Raman modes of the investigated species using $514$ nm laser excitation. a) bulk HOPG, b) SGN18 graphite powder, c) ultrasound sonicated exfoliated graphene, d) shear mixed exfoliated graphene, e) stirred exfoliated graphene. Solid color line represents Lorentzian-fits, grey lines denotes the decomposition of the 2D peaks. The dashed green line in case of the ultrasound sample points that the 2D peak can be fitted with one single Lorentzian as well.}
\label{fig:raman}
\end{figure}

Raman spectra of the examined samples are shown in Fig. \ref{fig:raman}. Namely, the D, G, and 2D Raman modes are presented. The D mode is associated with the presence of defects \cite{thomsen2000}. The 2D is its overtone. And the G (graphitic) mode is related to tangential motion of carbon atoms and it is the most pronounced in graphite. Solid lines represent Lorentzian fits. In most cases 2D lines are made up of 2 components, namely 2D$_1$ and 2D$_2$. In the case of ultrasound preparation the 2D feature can also be well fitted with one single Lorentzian. Parameters of the fitted Lorentzian curves are given in Table \ref{tab:raman}.

The Raman spectrum properties of graphite powder differ from HOPG. This is not an instrumental artifact, positions and widths of peaks in case of graphite strongly depends on morphology and grain size \cite{Wang1990}.

\begin{table}[h!] 
\begin{center}
  \begin{tabular}{cccccc} \hline
    $514$ nm                    & HOPG      & SGN18     & US    & SM   & ST \\ \hline
    $\nu_{\text{D}}$            & $1358.4$  & $1349.8$  & $1355.5$      & $1350.6$      & $1353.9$ \\
    $\Delta\nu_{\text{D}}$      & $18.6$    & $15.5$    & $20.0$        & $29.4$        & $14.2$ \\
    $\nu_{\text{G}}$            & $1583.3$  & $1579.0$  & $1583.3$      & $1581.6$      & $1582.2$ \\
    $\Delta\nu_{\text{G}}$      & $6.9$     & $8.1$     & $9.5$         & $10.6$        & $9.6$ \\
    $\nu_{\text{2D$_1$}}$       & $2688.4$  & $2686.2$  & $2696.4$      & $2692.8$      & $2692.4$ \\
    $\Delta\nu_{\text{2D$_1$}}$ & $21.4$    & $21.4$    & $25.7$        & $23.7$        & $23.9$ \\
    $\nu_{\text{2D$_2$}}$       & $2728.6$  & $2722.8$  & $2727.2$      & $2726.1$      & $2729.0$ \\
    $\Delta\nu_{\text{2D$_2$}}$ & $17.1$    & $19.6$    & $14.7$        & $17.4$        & $17.3$ \\
    $\nu_{\text{2D}}^{\ast}$    &           &           & $2714.6$      & & \\
    $\Delta\nu_{\text{2D}}^{\ast}$&         &           & $29.4$        & & \\ \hline
  \end{tabular}
  \caption{Parameters of the fitted Lorentzian curves for the D, G, and 2D Raman modes for a 514 nm excitation. $\nu$ denotes the position and $\Delta\nu$ the FWHM in cm$^{-1}$, $^{\ast}$ stands for single Lorentzian fit.}
  \label{tab:raman}
\end{center}
\end{table}

The D peak is less pronounced when ultrasound sonication or shear mixing was applied in case of exfoliated graphenes. The position of the D peak varies between the graphite powder and HOPG. According to mechanically exfoliated and CVD studies \cite{Ferrari2006,Malesevic2008} the D peak is expected at about $1350$ cm$^{-1}$ which is in a good agreement with our results. Both Ferrari and Das \cite{Das2008} agree that the intensity of the D peak for single layer material has to be negligible in order to assure a high quality of the material. The ultrasounded and shear mixed samples satisfies this criterion. The D peak is always present in wet chemically exfoliated graphenes \cite{Lotya2009,Englert2011} but its intensity is flake-size dependent \cite{Coleman2013}. The wet exfoliation according to the D peak intensity is far better in quality than for reduced GO samples \cite{Navarro2007,Stankovich2007,Zhang2010}.

All our FLG samples have a sharp G peak very close to HOPG (we remind that the starting material is SGN18). The width is about $\sim 2$ cm$^{-1}$ broader than graphite (both powder and bulk). Position of the G mode varies around $1580$ cm$^{-1}$ in good agreement with previous studies \cite{Ferrari2006,Das2008}. The G-line position also depends on the substrate and the number of layers. According to Ref. \cite{Chen2009}, the G peak position for the shear mixed and ultrasounded materials are very close to free-standing graphene.

The 2D peak for single layer graphene is expected to be a single, symmetric peak \cite{Ferrari2006}. The position of the peak is about $2700$ cm$^{-1}$ and shows a variation in the literature \cite{Ferrari2006,Das2008,Hernandez2008,Englert2009}. Width of the peak also varies in a wide scale from $15$ up to $40$ cm$^{-1}$. Variations can be explained with the effect of the substrate (samples on substrates always present a narrower peak) and the effect of preparations (strain, compressive forces may apply, and chemicals may remain). Our FLG samples show two components for the 2D line. The position of the lower 2D$_1$ peak agrees with previous single layer studies, thus this component is associated with single layer graphene sheets. The 2D$_2$ peak position is close to that of graphite. The presence of the 2D$_2$ mode can be interpreted as the presence of few layer sheets up to $4$ layers. The nominal width of the peaks suggest that we are dealing with single and few layer graphenes unlike in turbostratic graphite (in that case the width of 2D would be about $50$ cm$^{-1}$ \cite{Ferrari2006}). Bilayer graphene has a unique 2D peak made up of $4$ components \cite{Ferrari2006}, which is not present here. In case of the ultrasounded sample, the 2D peak can also be well fitted with one single Lorentzian with a position up to $2715$ cm$^{-1}$.

The amplitude ratios of 2D and G peaks are given in Table \ref{tab:raman_intratio}.

\begin{table}[h!] 
\centering
  \begin{tabular}{cccccc} \hline
    $514$ nm                        & HOPG   & SGN18  & US & SM & ST \\ \hline
    $I_{\text{2D}_1}/I_{\text{G}}$  & $0.21$ & $0.17$ & $0.21$     & $0.27$      & $0.37$  \\
    $I_{\text{2D}_2}/I_{\text{G}}$  & $0.42$ & $0.22$ & $0.22$     & $0.36$      & $0.25$  \\
    $I_{\text{2D}}/I_{\text{G}}$    & $0.63$ & $0.39$ & $0.43$     & $0.63$      & $0.62$  \\
    $I_{\text{2D}}^{\ast}/I_{\text{G}}$&     &        & $0.24$     &             &   \\ \hline
  \end{tabular}
  \label{tab:raman_intratio}
  \caption{Amplitude ratios of 2D and G peaks, $^{\ast}$ notes the single Lorentzian fit.}
\end{table}

Previous studies suggest that the number of layers can be extracted from this ratio \cite{Das2008,Ni2009,Tsurumi2013}. Several other effects, including the substrate (coupling-effect), the strain or compression, the way of preparation, the type and quality of the solvent and the wavelength of laser excitation also affects the 2D to G Raman signal ratio. Therefore the ratio of $I_{\text{2D}}/I_{\text{G}}$ has to be treated with care.
The ratio in case of mechanically exfoliated and CVD samples on substrates is greater than one. For free-standing graphene and wet exfoliated species always lower than one. Taking into account the previous considerations, wet exfoliated material is structurally more similar to free-standing graphene than the ones on substrates. The substrate may generate an extra damping for the G band phonons, which can lower the intensity of the G peak and change the ratio.

\subsection{Electron Spin Resonance spectroscopy}

\begin{figure}[h!]
\includegraphics[width=\linewidth]{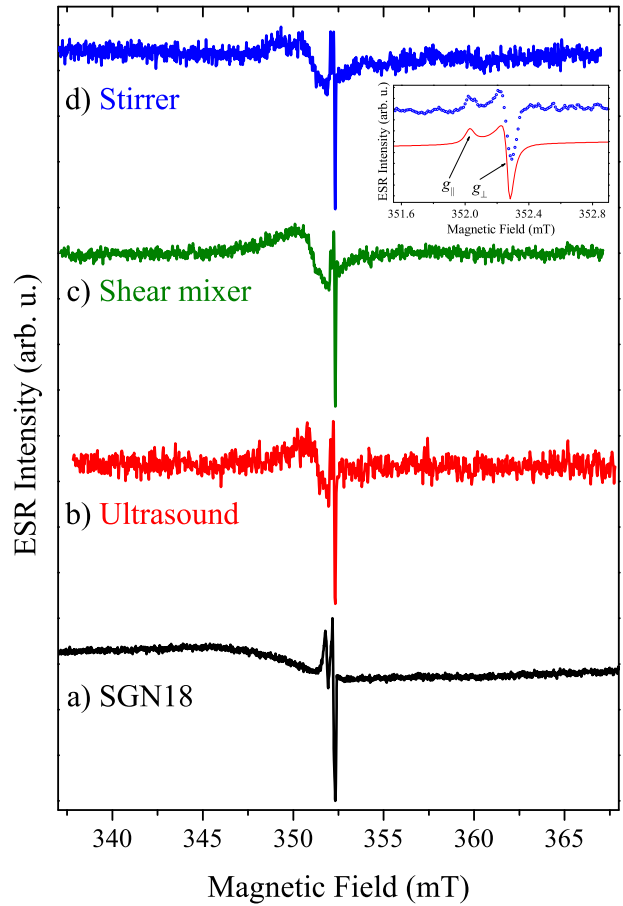}
\caption{ESR spectra of a) SGN18 graphite powder, b) ultrasounded, c) shear mixed, and d) stirred FLGs. The graphite powder has a broad line of about $12.2$ mTs as expected at a $g$-factor of $2.0148$. Ultrasounded FLG present a Lorentzian of $1.1$ mT linewidth at $g=2.0059$, the shear mixed present a $1.4$ mT at $g=2.0082$. The stirred material has a uniaxial anisotropic line with the width of $1.2$ mT at $g=2.0094$. The narrow uniaxial anisotropic line is coming from defects and dangling bonds in all cases. The inset shows the uniaxial $g$-factor simulated ESR lineshape for the narrow component in the stirrer prepared sample.}
\label{fig:esr}
\end{figure}

ESR spectra of the investigated materials are presented in Fig. \ref{fig:esr}. All samples (including the SGN18 starting material) show a narrow feature with a characteristic, uniaxial $g$-factor anisotropy lineshape shown in the inset of Fig. \ref{fig:esr}. This signal most probably comes from defects which are embedded in the sp$^2$ matrix of graphene, which may explain the uniaxial nature of the $g$-factor anisotropy.

The broader component for the SGN18 graphite sample has a characteristic $12$ mT ESR linewidth with a $g$-factor of $2.0148$ \cite{Huber2004,Galambos2009}. This line originates from conduction electrons present in graphite, the value of $g$-factor is the weighted average of the two crystalline directions ($B\parallel c$ and $B \perp c$) with $g$-factors of the two, which are present in HOPG \cite{Sercheli2002-1,Sercheli2002-2,Huber2004} with values of $2.0023$ and $2.05$. Here, $c$ is the direction perpendicular to the graphene sheets. The broader component has a $1.1-1.4$ mT linewidth for the three FLG samples with a $g$-factor slightly above the free-electron value $g_0=2.0023$. We tentatively assign this signal to a few layer graphene phase which is $p$-doped due to the solvent molecules. $p$-doping as in AsF$_5$ is known to give rise to similar signals with a $g>g_0$ \cite{Dresselhaus1981}. Ultrasounded and shear mixed materials present a single derivative Lorentzian peak with a width of $1.1$ mT and $1.4$ mT, respectively. The stirred sample displays a peak similar to that of graphite powder, but with a much narrower width of $1.2$ mT. The $g$-factor of FLG materials is between the free electron and the graphite powder. The most probable explanation for this is that single layer sheets are give a $g$-factor close to free electrons, but screened by the few layer sheets whose $g$-factor is closer to graphite. The sharp lines are associated with the defects and dangling bonds. In all materials the $g$-factor is above the free electrons $2.0023$, thus can be associated with $p$-type charge carriers. The spectra were simulated with derivative Lorentzian lineshapes whose parameters are given in Table 3.

\begin{table}[h!] 
\centering
  \begin{tabular}{ccccc} \hline
    Broad component    & SGN18  & US & SM & ST \\ \hline
    $g$           & $2.0148$ & $2.0059$   & $2.0082$    & $2.0094$ \\
     $\Delta B$ (mT) & $12.2$& $1.1$     & $1.4$      & $1.2$  \\ \hline
    Narrow component  & SGN18  & US & SM & ST \\ \hline
    $g$           & $2.0014$ & $2.0013$   & $2.0006$    & $2.0013$   \\
    $\Delta B$ (mT)  & $0.08$  & $0.04$      & $0.04$       & $0.04$ \\ \hline
  \end{tabular}
  \caption{$g$-factor, $\Delta B$ linewidth of the measured materials.}
  \label{tab:esr}
\end{table}

Previous study done by \'Ciri\'c \emph{et al} \cite{Ciric2009} on mechanically exfoliated graphene showed a $0.62$ mT wide peak with a $g$-factor of $2.0045$. On reduced GO \cite{Ciric2012} a $g$-factor of $2.0062$ and a width of $0.25$ mT was found. The solvothermally synthesized graphene \cite{Nafrady2014} shows a peak with a $g$-factor of $2.0044$ and a width of $0.04$ mT.
According to these studies wet exfoliated graphene species have a $g$-factor close to reduced graphite, but with a width close to mechanically exfoliated and solvothermally synthesized.

\subsection{Microwave resistance}

\begin{figure}[h!]
\includegraphics[width=\linewidth]{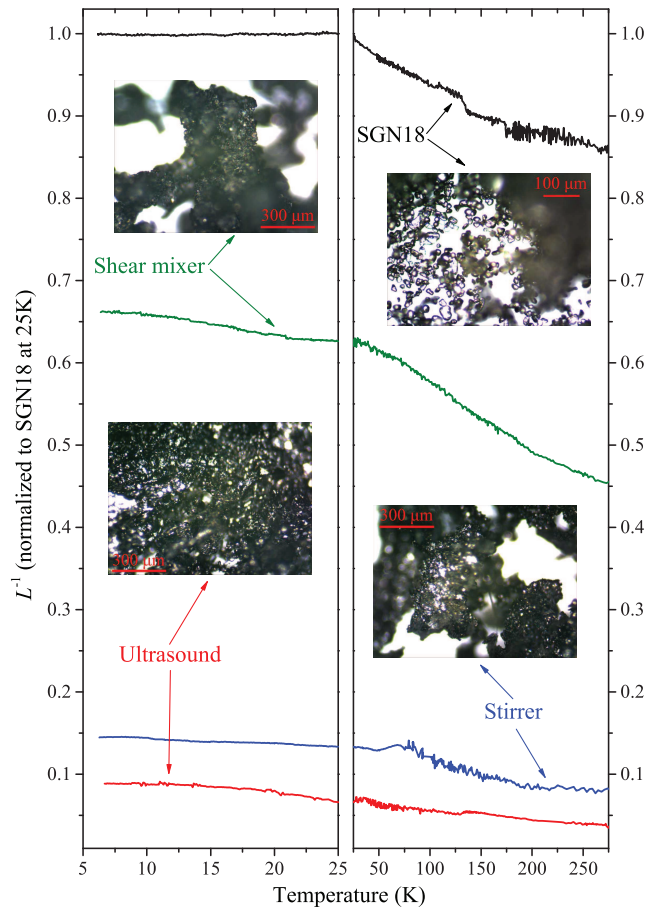}
\caption{Microwave resistance of FLGs compared to graphite. Insets are microscope images of the materials. Note the different scale for the SGN18 graphite sample. The different resistance of FLG species can be explained with the different grain size.}
\label{fig:mwcond}
\end{figure}

This results are presented in Fig. \ref{fig:mwcond}. This method is based on measuring the microwave loss due to the sample inside a microwave cavity. This contactless method is preferred when measuring resistance in powder samples, however the measured loss depends on the sample amount and morphology. It therefore provides accurate measurement of the \emph{relative} temperature dependent resistance, however it does not allow for a direct measurement of the resistivity. The resistance is proportional to the inverse of the microwave loss and it is normalized to that of SGN18 at $25$ K to get comparable results. Microscope images are presented as insets of Fig. \ref{fig:mwcond}. to demonstrate the difference in grain size.

All the measured materials have a semi-conducting behavior in the investigated temperature range. This behavior is usual to defective and inhomogeneous polycrystalline metals. The difference in the microwave loss in the different samples is primarily due to a difference in the grain size. The loss, $L$, is known to scale with the average grain size as $L = \pi B_0^2 \sigma R^5 / 5$, where $B_0$ is the amplitude of the magnetic field, $\sigma$ is the conductivity, and $R$ is the average radius of the grains \cite{Klein1993,Kitano2002}. The average grain size was obtained as about $3-5$ millimeters, $500$ $\mu$m, and $300$ microns for the ultrasounded, stirred and shear mixed samples, respectively, by analyzing the corresponding microscope images. The trend in the microwave loss between the different samples is thus found to follow the grain size.

\section{Conclusions}

We studied the vibrational, magnetic and transport properties of mechanically different post processed few layer graphene systems with Raman, ESR spectroscopy and microwave resistance measurements respectively. According to the results, processed treatment does affect the investigated properties of the material. From our results one can figure out that ultrasound treatment ends up with the best results in a meaning that this is the closest to a true single layer graphene, high quality and produced by a bulk synthesis method.  

\section{acknowledgement}
Work supported by the European Research Council Grant No. ERC-259374-Sylo. J. C. and S. R. acknowledge the DRS POINT-2014 Founding, P. Sz., B. N. and L. F. acknowledge the support of the Swiss National Science Foundation. The authors thank the Deutsche Forschungsgemeinschaft (DFG-SFB 953 "Synthetic Carbon Allotropes", Project A1) for financial support. The research leading to these results has received partial funding from the European Union Seventh Framework Programme under grant agreement no. 604391 Graphene Flagship.

\end{document}